**Complex Contagions: A Decade in Review**

Douglas Guilbeault, Joshua Becker and Damon Centola*

Annenberg School for Communication

Send correspondence to: dcentola@asc.upenn.edu

**Abstract**

Since the publication of "Complex Contagions and the Weakness of Long Ties" in 2007, complex contagions have been studied across an enormous variety of social domains. In reviewing this decade of research, we discuss recent advancements in applied studies of complex contagions, particularly in the domains of health, innovation diffusion, social media, and politics. We also discuss how these empirical studies have spurred complementary advancements in the theoretical modeling of contagions, which concern the effects of network topology on diffusion, as well as the effects of individual-level attributes and thresholds. In synthesizing these developments, we suggest three main directions for future research. The first concerns the study of how multiple contagions interact within the same network and across networks, in what may be called an ecology of contagions. The second concerns the study of how the structure of thresholds and their behavioral consequences can vary by individual and social context. The third area concerns the roles of diversity and homophily in the dynamics of complex contagion, including both diversity of demographic profiles among local peers, and the broader notion of structural diversity within a network. Throughout this discussion, we make an effort to highlight the theoretical and empirical opportunities that lie ahead.

## 1. Introduction

Most collective behaviors spread through social contact. From the emergence of social norms, to the adoption of technological innovations, to the growth of social movements, social networks are the pathways along which these "social contagions" propagate. Studies of diffusion dynamics have demonstrated that the structure (or topology) of a social network can have significant consequences for the patterns of collective behavior that will emerge.

Over the last forty-five years, questions about how the structure of social networks affects the dynamics of diffusion have been of increasing interest to social scientists. Granovetter's (1973) "Strength of Weak Ties" study ushered in an era of unprecedented interest in how network dynamics, and in particular diffusion on networks, affect every aspect of social life, from the organization of social movements to school segregation to immigration. Granovetter's study showed that "weak ties" between casual acquaintances can be much more effective at promoting diffusion and social integration than "strong ties" between close friends. This is because although casual friendships are relationally weak, they are more likely to be formed between socially distant actors with few network "neighbors" in common. These "long ties" between otherwise distant nodes provide access to new information and greatly increase the rate at which information propagates, despite the relational weakness of the tie as a conduit.

In the last two decades, the explosion of network science across disciplines such as physics, biology, and computer science has produced many important advances for understanding how the structure of social networks affect the dynamics of diffusion. The full impact of Granovetter's original insight was not realized until Watts and Strogatz's (1998) "small world" model demonstrated that bridge ties connecting otherwise distant nodes can dramatically increase the rate of propagation across a network by creating "shortcuts" between remote clusters. Introducing "long ties" into a network can give even highly clustered networks the "degrees of separation" characteristic of a small world. This model of network dynamics has had a tremendous impact on fields as diverse as computer science, physics, epidemiology, sociology, and political science.

Building on the idea of pathogenic contagions, this research combines the diverse domains of ideas, information, behaviors and diseases into the generic concept of a universal contagion. The attractive implication is that the mathematical tools developed by epidemiologists for studying the spread of disease can be generically used to study the dynamics of social, cultural and political change. In particular, the properties of social networks that have been shown to accelerate the spreading dynamics of disease diffusion–such as small world topologies, weak ties, and scale-free degree distributions –can also be used to make inferences about the role of networks in the domains of social and political behavior. Regardless of whether a given contagion is a prophylactic measure to prevent HIV infection or the HIV infection itself, Granovetter's groundbreaking claim was that "whatever is to be diffused can reach a larger number of people, and traverse a greater social distance, when passed through weak ties rather than strong" (Granovetter 1973: 1366).

However, while this theory is useful for understanding the rapid spread of HIV infections through networks of weak ties, it has not shed light on the remarkable failure of these same networks to spread prophylactic measures for preventing HIV (Coates et al. 2008). The reason for this disturbing asymmetry between the spread of infectious diseases and the diffusion of preventative measures is that infectious diseases are typically *simple contagions* –i.e., contagions for which a single activated source can be sufficient for transmission –while preventive measures are typically *complex contagions* –i.e., behaviors, beliefs, or attitudes for which transmission requires contact with multiple sources of activation. While repeated contact with the same person can increase the likelihood of transmitting a simple contagion, the transmission of complex contagions requires reinforcement from several contacts. Any social contagion that is costly, difficult, or unfamiliar is likely to be a complex contagion, requiring social reinforcement to spread.

The primary consequence of the distinction between simple and complex contagions for diffusion through social networks is that as "worlds" become very small, the speed of simple contagions increases, while complex contagions become harder to spread. As Centola and Macy write,

> For simple contagions, too much clustering means too few long ties, which slows down cascades. For complex conta-
gions, too little clustering means too few wide bridges, which
not only slows down cascades but can prevent them entirely.
(2007: 723)

Centola and Macy (2007) identify several reasons why contagions
may be complex, including the need for social legitimation, the need
for credibility, or the complementarity of a behavior. For instance, a
contagion might be complex due to externalities, in which the value
of the contagion increases with the number of adopters. The value of
a communication technology such as a fax machine rests heavily on
the number of people who use it. When only one person has a fax
machine, it holds no value. A single contact with someone who has a
fax machine provides little reason for someone else to adopt it. Even
if the adopter provides repeated signals, a single person alone cannot
do much to increase the complementary value of the fax machine.
However, if a potential adopter comes into contact with several inde-
pendent sources who have all adopted fax machines, the complemen-
tary value of the technology increases. After exposure to a sufficient
number of reinforcing contacts, a person with no inherent interest in
fax machines can be convinced that it is a necessary investment.

    A different kind of reason why a contagion might be complex
is due to uncertainty. For instance, physicians are often resistant to
adopting new medical technologies for fear of placing themselves at
risk of acting outside of accepted protocols. Early studies on adoption
patterns among physicians found that physicians were unlikely to
adopt a new medical technology, even though it had been formally
approved and was expected to be very effective, until they observed
several of their colleagues using it (Coleman et al., 1966). For similar
reasons, complexity in diffusion can also be a result of normative
pressures. This is often the case with the diffusion of managerial prac-
tices among elite firms. Because the choice of corporate governance
strategy can impact the reputation of a firm, the adoption of new prac-
tices is often dependent upon social reinforcement from competing
firms within the same industry. Corporate boards concerned about the
risk of social sanction are often unwilling to adopt new managerial
practices until they have already seen them adopted by several peer
institutions (Davis and Greve 1997).

In the last decade, the literature on complex contagions has rapidly evolved both empirically and theoretically. In this review, we discuss recent developments across four empirical domains: health, innovation, social media, and politics. Each domain adds new complexities to our understanding of how contagions emerge, spread, and evolve, as well as how they undergird fundamental social processes. We also discuss how these empirical studies have spurred complementary advancements in the theoretical modeling of contagions, which concern the effects of network topology on diffusion, as well as the effects of variation in threshold dynamics. Importantly, while the empirical studies reviewed in this paper complement existing theoretical work, they also present many opportunities for new theoretical extensions. We suggest three main directions for future development of research on complex contagions. The first concerns the study of how multiple contagions interact within the same network and across networks, in what may be called an ecology of complex contagions. The second concerns the study of how the structure of thresholds and their behavioral consequences can vary by individual and social context. The third area concerns the recent discovery of diversity as a causal variable in diffusion, where diversity can refer either to the diversity of demographic profiles among local peers, or to the broader structural diversity that local peers are situated within. Throughout, we take effort to anticipate the theoretical and empirical challenges that may lie ahead.

## 2. Empirical Advances

### 2.1. Applications to Health

For the past few decades, the study of public health has concerned not only biological contagions, but also social contagions concerning health behaviors: e.g. medication, vaccines, exercise, and the ideologies related to each (Christakis and Fowler 2012). It has been found that simple contagions do not adequately capture the network dynamics that govern the diffusion of health behaviors (Centola and Macy 2007; Centola et al. 2007; Centola 2010, 2011). Social health behaviors often require reinforcement from peers, and they are strongly influenced by cultural practices and group norms.

The Framingham Heart Study suggested that obesity spread socially through a densely interconnected network of 12,067 people, assessed from 1971 to 2003 (Christakis and Fowler 2007). However, this study posited that either biological or normative mechanisms might play a role in the diffusion process, where each mechanism would be expected to yield very different diffusion dynamics.

A clearer hypothesis came from a follow-up study examining the spread of smoking behavior (Christakis and Fowler 2008). This study found evidence that the likelihood a smoker will quit depends on their exposure to multiple contacts, in part because smoking is often explicitly social and thus shaped by the dynamics of social norms. The role of complexity in smoking behavior (and cessation) has been supported by a more recent study using data from the National Longitudinal Study of Adolescent Health, which simulated the complex contagion dynamics of smoking under conditions where smokers can revert to smoking after quitting (Kuhlman et al. 2011). By examining peer interactions over QuitNet – a social media platform for smokers attempting to quit – it was found that smokers were more likely to abstain if exposed to reinforcing contact from several abstinent users (Myeni et al. 2015). Kuhlman et al. (2011b) discuss how the diffusion of smoking behavior is filtered by both pro- and anti-smoking norms. This insight into the complexity of the quitting process helps to refine earlier models of smoking diffusion, in which threshold outcomes are represented by the binary decision to adopt without consideration of countervailing influences from non-adopters. Norms empower people to exert different kinds of influence – eg. for and against behavior – which amplifies the role of complexity in situations where non-adopters exhibit countervailing influences.

Exercise has similarly been found to exhibit the dynamics of complexity when peers influence each other to adopt new exercise behaviors. The characteristics of peers play an important role in influence dynamics, as both homophily and diversity have been shown to amplify the impact of reinforcing signals on the likelihood of behavior change. Centola (2011) demonstrated a direct causal relationship between homophily and the diffusion of complex contagions, indicating that the effects of social reinforcement were much stronger when individuals shared a few key health characteristics in common. Further, Centola and van de Rijt (2014) showed that social selection among "health buddy" peers in a fitness program lead to connections among peers who were homophilous on the same key health characteristics: gender, age, and BMI. Aral and Nicolaides (2017) elaborate in showing that social reinforcement from similar peers is strengthened when those peers come from different social groups, highlighting the value of structural diversity in the dynamics of complexity. Another recent study of exercise behavior used an online intervention to demonstrate that exposure to social influence from a reinforcing group of anonymous online "health buddies" could directly increase participants' levels of offline exercise activity (Zhang et al. 2016).

An interesting twist in the relationship between complexity and health came from a series of studies which showed how clustered networks that facilitate the spread of social norms (e.g., anti-vaccination behavior) can thereby make populations susceptible to epidemic outbreaks of simple contagions (e.g., such as the measles) (Salathe and Bonhoeffer 2008; Campbell and Salathe 2013). These studies model the diffusion of anti-vaccine attitudes as a complex contagion that pulls people into echo chambers that amplify the likelihood of disease outbreak in the overall population. This work points to a vital direction for future research into how health behaviors and attitudes toward health interact in a broader, multi-layered network of both complex contagions and disease diffusion.

Moreover, there are even some surprising instances where biological pathogens may also be complex. Infectious diseases are complex in situations where patients suffer simultaneous "co-infections" from multiple pathogens. In these cases, each disease increases a patient's susceptibility to the other one, making it more likely that both infections will take in hold in a patient. For instance, infection with the influenza virus can increase the likelihood of coinfection with other respiratory diseases, such as the *Streptoccocus penumoniae* bacterium (a leading cause of pneumonia). Each one creates susceptibility to the other, increasing the likelihood that joint exposure will lead a patient to become infected with both.

While a single virus can efficiently use weak ties to spread across a network, several viruses from different sources cannot be so easily transmitted the same way. For these kinds of illnesses, clustered social networks significantly increase the likelihood that individuals who are exposed to complementary infections, such as pneumonia and flu, or syphilis and HIV, will spread reinforcing co-infections.

Contrary to most epidemiological intuitions, in random networks incidence rates of "complex synergistic co-infections" typically drop to zero, while clustered social networks are surprisingly vulnerable to epidemic outbreaks (Hébert-Dufresne and Althouse 2015).

## 2.2. Diffusion of Innovations

Economists, marketers, and organizational theorists have long been interested in how technological innovations diffuse through a population. Bass (1969) developed one of the first influential models of innovation diffusion, where technological adoption was understood as a simple contagion. As the uptake of innovations came to be viewed as inseparable from social networks, Schelling (1973) started to formulate a threshold-based model of innovation adoption based on the influence of multiple peers. It has since been found that complex contagions characterize the diffusion of technologies in multiple areas of social life.

A number of controlled experiments illustrate that innovations diffuse through populations as complex contagions. Bandiera and Rasul (2006) showed how farmers in Mozambique were more likely to adopt a new kind of crop if they had a higher number of network neighbours who had adopted. Oster and Thorton (2012) show that the adoption of menstrual cups in women depends on influence from multiple peers, because of the transference of technology-relevant knowledge. Based on these findings, Beaman et al. (2015) used complex contagion models to design seeding strategies for the distribution of pit planting in Malawi. Pit planting is a traditional West African technology which is largely unknown in Malawi, and it has the potential to significantly improve maize yields in arid areas of rural Africa. Beaman et al. compared the seeding strategies recommended by complex contagion models to a benchmark treatment where village leaders used local knowledge to select seeds. Seeding, in this experiment, involved training specific people in each village on how to use pit planting, given evidence that trained adopters of a technology are most effective at distributing new technologies (Banerjee et al. 2013). 200 different villages were randomly treated with seeding strategies from either complex contagion models or traditional approaches based on local expertise. They found that seeding strategies informed by complex contagion models increased adoption more than relying on extension workers to choose seeds. Further, Beaman et al. observe no diffusion of pit planting in 45% of the benchmark villages after 3 years. In villages where seeds were selected using the complex contagion model, there was a 56% greater likelihood of uptake in that village.

Complex contagions have also been shown to characterize the diffusion of software innovations. Karsai et al. (2014) examined the uptake of Skype – the world's largest Voice over internet protocol service – from September 2003 to March 2011. They find that the probability of adoption via social influence is proportional to the fraction of adopting neighbours. Interestingly, they find that while adoption behaves like a complex contagion process, termination of the service occurs spontaneously, without any observable cascade effects. These results suggest that there may be an asymmetry in the dynamics of adoption (which are socially driven) versus the dynamics of termination (which may depend on non-social factors).

Ugander et al. (2012) also observe complex contagion dynamics in the initial growth of Facebook, which now has over a billion users worldwide. Facebook initially grew through peer recruitment over email. The results showed a complex diffusion process, in which people were more likely to adopt Facebook if they received requests from multiple friends, especially if these friends belonged to separate network components. This finding on the value of structural diversity for amplifying reinforcing signals for adoption suggests interesting new theoretical directions for research on the connections between homophily, diversity, and complexity (see Sect. 4.3).

A parallel stream of research has focused on the role of mass-media marketing in spreading the complex diffusion of innovations. Toole et al. (2012) show that while mass media served to measurably increase the adoption of Twitter, peer to peer social influence mechanisms still account for the lion's share of the adoption patterns that were observed, where local reinforcement played a major role in individuals' decisions to adopt Twitter. So much so, that the online microblogging platform exhibited strong spatial diffusion patterns in its initial growth, as it spread through densely clustered networks of peer reinforcement. Similar findings are echoed by Banerjee et al. (2013). These studies suggest that the local peer influence dynamics of complexity can initiate global cascades in the adoption of innovations. For marketing to propel the diffusion of new technologies, mass-media strategies need to account for how messages are dynamically filtered by social networks (Berger 2013). Evidence suggests that advertising campaigns initially diffuse like simple contagions with the first media broadcast, but diffuse more like complex contagion once they begin spread through social networks (Bakshy et al. 2012a). The interaction between mass-media diffusion and social influence in the adoption of technology (particularly complementary technologies) suggests that the complexity of a diffusion process is determined in part by interactions across several scales of a population.

The study of innovation diffusion is expanding in response to a novel kind of complexity introduced by technologies themselves. A new direction for future research concerns the role that social media technologies play in shaping the evolution of other contagions, once the social media technologies themselves are adopted. Due to their explicitly complementary design, social media technologies, including Facebook, Twitter, and Skype, all exhibited the dynamics of complexity in their diffusion, spreading most effectively through networks of peer reinforcement. Once these technologies diffuse, they allow individuals to grow larger networks that communicate at much faster rates than were previously possible in word-of-mouth exchanges. Thus, in addition to the spread of social media technologies, the domain of social media itself has become its own space for studying the complex dynamics of the diffusion of collective behavior.

### 2.3. Social Media

Social media has significantly shaped and, in some cases, augmented the diversity of complex contagions that can spread, the speed at which they can spread, and the overall size of the populations they are able to reach via global cascades (Borge-Holthoefer et al. 2013). Kooti et al. (2012) show that one of the first methods for retweeting was established as the successor of various competing complex contagions, in an ecology of possible conventions. Barash (2011) and Weng et al. (2013) find that most tweets spread via complex contagions in retweet networks. This finding reappears with Harrigan et al. (2012) who show that tweets are more likely to diffuse through retweeting within clustered communities, where twitter users are able to observe their friends retweeting the same message. Complex contagions are observed across other platforms as well. Photo-tagging in Flickr exhibits the hallmarks of diffusion via influence from multiple peers (Barash 2011). A recent massive-scale randomized experiment over Facebook showed that user-generated stories diffused like complex contagions (Bakshy et al. 2012b). Meanwhile, social media websites gather an unprecedented amount of data on communication flows, permitting novel insights into how complex contagions emerge and operate.

One of the most interesting findings of social media research is that the content of a contagion matters for whether it behaves in a complex manner. Wu et al. (2011) show that the modality of information that structures a contagion influences its lifespan: viral videos long outlive their textual counterparts. Romero et al. (2011) find that there are distinct contagion dynamics for different kinds of hashtag. Political hashtags are found to behave like complex contagions, where exposure to multiple people using the hashtag is strongly correlated with use. But hashtags based on idioms or memes, by contrast, behave like simple contagions. Barash and Kelly (2012) and Fink et al. (2016) replicate this finding by showing that political hashtags behave like complex contagions, whereas news-based hashtags, broadcast by mass media, spread like simple contagions.

Using the massively multiplayer virtual world of Second Life, Bakshy et al. (2009) uncover complex contagions in the exchange of user-created content. Specifically, they focus on the spread of conventionalized avatar gestures constructed by players, which can only spread through peer to peer sharing mechanisms. Bakshy et al. unveil subtle interactions between user degree and diffusion: users who are most effective at initiating cascades of gestures do not have the highest degree; rather, they collect rare gestures that other users are more likely to adopt. This result points to uncharted territory in complex contagions research, relating to how the quality or style of a contagion influences its likelihood of spreading via social influence.

Undoubtedly, the source of complexity in these online dynamics of spreading behavior lies partly in the sociological significance that the content of an online contagion holds. For instance, Romero et al. suggest that political hashtags, such as #TCOT (which stands for "Top Conservatives on Twitter") and #HCR (which stands for "Health Care Reform") were "riskier to use than conversational idioms…since they involve publicly aligning yourself with a position that might alienate you from others in your social circle." (2011: 3). Thus, the authors found "that hashtags on politically controversial topics are particularly persistent, with repeated exposures continuing to have unusually large marginal effects on adoption" (3).

It is also likely that the level of complexity in diffusion depends, in part, on the design of interfaces and the kinds of sociological processes that platforms facilitate. Readymade communication buttons – such as the 'share' button on Facebook or the 'retweet' button on Twitter – automatically enable the spread of information as a simple contagion. However, State and Adamic (2015) show how simple contagions do not account for the spread of digital artifacts that require more effort to construct. Using a dataset of over 3 million users, they show that the adoption of new conventions for profile pictures are best described as complex contagions. They argue that the difference pertains to the amount of effort it takes to adopt the behavior: certain informational contagions behave in a simple manner because it takes no time to click and share after one exposure. But when a contagion requires more effort, such as manually changing a profile picture, users require evidence that several of their peers have expended the energy for the contagion, thereby justifying its weight in terms of social capital.

Conversely, platform design can also prevent complex contagions from emerging and spreading by constraining the ability for people to perceive and share potential contagions (Bakshy et. al. 2012a, 2012b; Hodas and Lerman 2012; Gomez-Rodriguez et al. 2014). Doerr et al. (2012) find that, over the social news aggregator *Digg*, users do not seem to preferentially share the content of their peers. This result is likely to be specific to the *Digg* environment, because the culture of the platform is based on sharing news that your friends do not already know. Studies of social media thus reveal how environmental design alters the capacity for diffusion by shaping the salience of peer behaviors and the culture of interaction altogether.

Going forward, social media environments are likely to serve as a powerful tool for studying complex contagions experimentally. Centola (2010, 2011, 2013a) developed a method for designing social media platforms that embed participants into engineered social networks, which allow researchers to test the effects of network topology and other variables on the dynamics of social diffusion. In a less controlled study, Kramer et al. (2014) modified the newsfeeds of Facebook users to examine emotional contagion. For some users, they reduced the amount of positive content, whereas for other users, they reduced the amount of negative content. As a result, they were able to systematically alter the emotional content of users' posts. While this study could not eliminate endogeneity within user networks, the randomization of messages allowed for suggestive experimental results on the ways that social exposure to messages influence user behavior.[1]

Another related approach to experimentation on social media comes from the advent of experimental methods that use algorithmically controlled social media accounts called bots to manipulate users' experiences (Krafft et al. 2016). Mønsted et al. (2017) released a network of bots into twitter and tested whether they could prompt the uptake of specific hashtags. They show that bots can initiate the uptake of new hashtags and that these hashtags spread as complex contagions, whereby the probability of using the new hashtag drastically increased if multiple bots and users were seen using it.

## 2.4. Politics

Political processes have been a longstanding topic of interest for threshold-based contagion models. Granovetter's (1973, 1978) original threshold model of collective action gave special attention to the start-up problem for political protests and riots. He observed that individuals have different degrees of willingness (i.e. thresholds) to participate in a riot, where their willingness is dependent on how many of their neighbours they observe participating in the riot. Granovetter observed that riots can emerge as a result of cascades, where a subset of instigator individuals with low thresholds trigger the spread of rioting. The first efforts to describe the emergence of social movements with agent-based modeling maintained that population diversity was essential for getting a movement off the ground. Without long ties connecting communities, it was thought that social movements would not be able to diffuse through a population and reach critical mass.

---

[1] Kramer et al.'s study also raised the important point about the ethics of experimentation on social media. While previous social media studies using experimentally designed social platforms (Salganik et al. 2006, Centola 2010, 2011, Centola and Baronchelli 2015) enrolled subjects into their online platform with an explicit process of informed consent, Kramer et al.'s study on Facebook used existing networks of peers without their explicit consent. It is an important topic of ongoing discussion how to properly use existing peer networks, such as Facebook and Twitter, to conduct experiments that manipulate user behavior.

More recent models extend the study of diversity in political processes by emphasizing the supporting role of homophily during the growth phase of social movements. Centola (2013b) argues that because social movements involve risky and costly forms of deviant behavior, people require reinforcement from multiple peers to participate, where homophily is useful for establishing a critical mass of likeminded peers.

Again, this raises an interesting connection between diversity and homophily. For organizing a critical mass, dense, homophilous communities are necessary for getting social movements off the ground because like-mindedness facilitates group solidarity, which may be necessary to withstand the normative backlash that comes from deviant behavior. On these grounds, Centola designed an agent-based model to show that weak ties hinder the spread of social movements by increasing exposure to counter-norm pressures, while also reducing the group transitivity needed to reinforce group interests. Homophily and clustering thus reinforce one another. However, once homophilous networks gain enough local reinforcement, they can create a critical mass that allows the movement to achieve sufficient salience in the whole population and to expand to diverse communities through the aid of mass media.

In an empirical study of the effects of communication networks on mobilization, Hassanpour (2017) explored the spread of armed conflict as a complex contagion in Damascus, Syria. On November 29[th], 2012, Internet and cellular communications were shut down all across Syria for over a day. The shutdown, according to Hassanpour, resulted in the loss of communication with long ties to individuals across the city. At the same time, the shutdown immediately preceded an unprecedented increase in the diffusion of armed conflict throughout the city. Using a geolocated dataset of daily conflict locations in Damascus, Hassanpour uncovers signs that the likelihood of conflict in a region was influenced by whether there had been conflict in multiple neighboring regions. Hassanpour suggests that this indicates the spread of conflict as a complex contagion, which was allowed to emerge when long ties were broken and interaction within local clusters became the strongest determiners of armed conflict.

In other results, González-Bailón et al. (2011) shows that protest recruitment in Spain, 2011, diffused over Twitter as a complex contagion via peer influence. González-Bailón et al. (2011) use k-core decomposition to show that the users who are in the core of the network were most effective at initiating cascades of recruitment. In a complementary study, Steinhert-Threlkeld (2017) offered evidence that users in the periphery of social media networks can also trigger global cascades. These studies suggest that social media can influence the rise and spread of political complex contagions that inspire on-the-ground political action.

Other recent empirical work has uncovered complex contagions within a wide range of political processes, including campaign donations (Traag 2016), grassroots mobilization (Parkinson 2013; Parigi and Gong 2014), petition signing (Yasseri et al. 2014), social control (Siegel 2009, 2011), institutional change (Dellaposta et al. 2017), and administrative management in both rural (Catlaw and Stout 2016) and urban settings (Pan et al. 2013).

Barash (2011) developed a unique set of measures for characterizing the lifespan of political contagions over social media. A complex contagion begins by saturating a locally clustered community. Once saturation is reached, the rate of propagation for the contagion decelerates, as the number of potential adopters decreases. If the saturated community has sufficiently wide bridges to other communities, Barash (2011) argues that it is possible for a contagion to travel from one community to the next. Diffusion between communities can create a detectable temporal signature, because as a contagion enters a new community, its rate of propagation rapidly increases with the availability of new adopters. Barash explains how changes in the rate of complex propagation can provide a measure for whether a contagion is ramping up for a global cascade, hinting toward the possibility of detecting global cascades, prior to their emergence.

Based on the work of Barash et al. (2012), Fink et al. (2016)

developed a number of measures for characterizing the spread of political hashtags as complex contagions. These measures include: *peakedness, commitment, concentration,* and *cohesion.* Peakedness concerns the duration of global activity associated with a contagion, where a peak refers to a day-long period of usage when the average mentions per day is more than two standard deviations away from the average mentions in the preceding days. Peakedness is closely related to burstiness, which has been shown to play an important role in threshold-based cascade dynamics (Takaguchi et al. 2013). Commitment refers to the number of people who sustain the life of a complex contagion, even though they endure social costs by not conforming to surrounding norms. Concentration simply refers to the proportion of people using a hashtag during a given time period. And cohesion refers to the network density over the subgraph of all users engaged in a particular contagious phenomena. The authors make use of the idea that complex contagions are incubated in locally dense communities before they colonize other communities via sparse connections.

Using these measures, researchers have made a number of valuable observations. Fink et al. (2016) apply these measures to the study of political hashtags in Nigeria. In their sample, they find political hashtags consistently arise with a small proportion of instigators (roughly 20%) who are densely connected, and that almost 60% of late adopters for political hashtags had 2 or more previous adopter friends. News hashtags, by contrast, are first propagated by largely unconnected instigators who constitute between 50% and 90% of the network, where less than 10% of adopters had 2 or more previous adopter friends. Consistent with Romero et al. (2011), the authors suggest that political hashtags require influence from multiple peers because they have higher social costs, especially in countries like Nigeria where surveillance by governments and extremists groups looms over users. Compared to other hashtags, the researchers also find that hashtags related to social movements have a higher density of ties among early adopters, consistent with the argument that political movements require a coalition of homophilous, densely connected users (Centola 2013b). Fink et. al further illustrate that it is possible to map the virality of political hashtags using Barash's measures for the temporal signatures of diffusion. They show how the #bring-backourgirls hashtag went viral shortly after a period of decreased usage among early adopters, which indicated saturation in a local community prior to the spread of the contagion to other communities.

In a related paper, Barash and Kelly (2012) use the same measures to model the spread of complex contagions over Russian Twitter, a significantly different cultural setting. Yet again, these researchers find that politically-salient hashtags diffuse like complex contagions where news hashtags from mass media do not. Importantly, this analysis shows how the heterogeneous distribution of adoption thresholds is critical for understanding political contagions. They find that engagement in political issues is non-uniform across the population, and different communities have distinct patterns of engagement and adoption, based on how the community relates to the content captured by the hashtag. Users belonging to groups that oppose the political regime engage with controversial topics over a long period, as a committed minority. Contrary to expectations, they find that when hot button issues relevant to the opposition make it into the mainstream, they are much more likely to sustain global saturation, even amongst pro-government users. These results suggest that reinforcement dynamics can drive the spread of politically-salient content over social media.

The diffusion of political contagions online interacts with both the structure of the sub-communities that they reach and the group identities that they activate. In March of 2013, 3 million Facebook users changed their profile picture to an 'equals sign' to express support of same-sex marriage. Consistent with earlier work, State and Adamic (2015) found that the equals sign profile picture spread as a complex contagion. Their data suggests that mass media created only about 58,000 spontaneous adopters, while roughly 106 million users adopted based on peer exposures. They find that it took, on average, exposures to 8 different peer adopters for a person to adopt. When examining this threshold, the authors uncover intricate dependencies between the identity of a user and their willingness to adopt. Users were more likely to be exposed and thereby more likely to adopt if they were female, liberal, non-heterosexual, and between the ages of 25-34. These findings suggest that thresholds for adopting contagions are modulated by online identity signaling regarding political values and beliefs. Similar, smaller scale studies of behavior on Facebook find that a user's demographic characteristics do not determine their influence in generating cascades, but instead most cascades rely on multiple users to trigger spreading (Sun et al. 2009).

The study of political contagions – offline and online – reveals a number of subtleties in how thresholds operate in sociological contexts. Political identity is a driving motivation for behavior change, suggesting that homophily and clustering in social networks can be essential for incubating the early growth of a political behavior over social media. Furthermore, gender, race, and religion are also strong predictors of whether someone will be exposed and receptive to a political contagion. A recent study by Traag (2016) shows that campaign donations diffuse as complex contagions, but the findings here emphasize the value of diversity. The growth of support for a candidate increases when people are exposed to donors from separate communities, particularly if those donors supported the opposite party. Diversity can thus complement homophily when it signals wider support for a candidate, and thereby increases the likelihood that the candidate will be more effective for achieving bi-partisan goals. The details are subtle, however, since there are also situations where diverse support for a candidate might signal mixed allegiances and comprise the candidate's party loyalty. The complementary roles of homophily and diversity in supporting complexity depend upon the content of the political messages that are used and the identities that they activate.

## 3. Theoretical Advances

Recent research into the formal model of complex contagions has explored two general directions. The first direction investigates how complex contagions spread within large networks of varying topologies. To date, researchers have examined threshold-based contagion models within power-law (Barash et al. 2012), locally tree-like (Gleeson and Cahalane 2007), degree-correlated (Gleeson 2008; Dodds and Payne 2009), directed (Gai and Kapadia 2010), weighted (Hurd and Gleeson 2013), small-world (Centola et al. 2007), modular (Galstyan and Cohen 2007), clustered (Ikeda et al. 2010; Hackett et al. 2011), temporal (Karimi et al. 2013; Backlund et al. 2014), multiplex (Brummitt et al. 2012; Yağan et al. 2012, Lee et al. 2014), and interdependent lattice networks (Shu et al. 2017). Researchers have used different topologies to simulate how external factors like mass media influence cascade dynamics (Bassett et al. 2012), and how topologies influence percolation processes (Zhao et al. 2013). A pivotal theoretical finding is that complex contagions require a critical mass of infected nodes to initiate global cascades, and it has been shown that critical mass dynamics depend in sensitive ways on network topology and the distribution of node degree and adoption thresholds (Barash 2011). There has also been efforts to provide analytic proofs for the global dynamics of complex contagions (Barash 2011; O'Sullivan 2015). At the cutting-edge is research into how complex contagions spread in coevolving, coupled, time-varying, and multi-layered networks (Takaguchi et al. 2013; Pastor-Satorras et al. 2015).

The second major direction in theoretical complex contagion research concerns mechanisms of diffusion at the node-level, concerning individual attributes and thresholds. Wang et al. (2015a, 2015b) propose a contagion model which shows that the final adoption size of the network is constrained by the memory capacities of agents and the distribution of adoption thresholds. Perez-Reche et al. (2011) simulate complex contagion dynamics with synergistic effects among neighbors, and McCullen et al. (2013) structure the motivation for an agent to adopt a behavior as a combination of personal preference, the average of the states of one's neighbours, and the global average. Dodds et al. (2013) attempt to explicitly encode sociological processes into their models by building agents with a preference for imitation but an aversion for complete conformity. Melnik et al. (2013) model *multi-stage* complex contagions, in which agents can assume different levels of personal involvement in propagating the contagion, at different times in their lifecycle. They find that multi-stage contagions can create multiple parallel cascades that drive each other, and that both high-stage and low-stage influencers can trigger global cascades. Huang et al. (2016) build agents with a persuasiveness threshold which determines their ability to initiate adoption. This new parameter can cause networks to become more vulnerable to global cascades, especially heterogeneous networks. Further incorporating sociological considerations, Ruan et al. (2015) simulated how conservativeness among nodes - i.e. the reluctance to adopt new norms - interacts with cascades caused by spontaneous adopters.

The latest theoretical developments have informed research on how to design network interventions and seeding strategies to stop the spread of harmful complex contagions (Kuhlman et al. 2015). Such interventions are based on the use of oppositional nodes that are permanently unwilling to adopt a behavior, regardless of peer influence. Kuhlman et al. (2015) offers two heuristics for using seeding methods to determine critical nodes for inhibiting the spread of complex contagions. The first heuristic is to select the nodes with the highest degree, and the second heuristic is to select nodes from the 20 core, determined by k-core decomposition. They show how the second heuristic is more effective for initiating and preventing global cascades, because selecting from the 20 core increases the likelihood that nodes are adjacent and thereby capable of reinforcing each other's influence. Centola (2009) shows that similar ideas can be used to evaluate the tolerance of networks against error and attack. Albert et al. (2000) showed that scale-free networks are robust against network failures, defined in terms of the inability to diffuse simple contagions. When it comes to diffusing complex contagions, Centola shows that scale-free networks are much less robust than exponential networks. Thus, moving from simple to complex contagions changes the robustness properties of scale-free networks. Building on this work, Blume et al. (2011) investigate which topologies are more susceptible to what they call *cascading failures*, which refers to the outbreak of negative complex contagions that are harmful for social networks. Siegel (2009, 2011) shows how these developments can inform models for repressing social movements and performing crowd control on the behalf of governments.

While early models of diffusion consider individual contagions as independent and spreading in isolation, a number of studies have begun to investigate evolutionary dynamics among multiple complex contagions. Myers and Leskovec (2013) develop a statistical model wherein competing contagions decrease one another's probability of spreading, while cooperating contagions help each other in being adopted throughout the network. They evaluate their model with 18,000 contagions simultaneously spreading through the Twittersphere, and they find that interactions between contagions can shape spreading probability by 71% on average. Jie et al. (2016) construct a similar model to simulate competing rumor contagions in a homogenous network. Empirical evidence is accumulating that multiple contagions frequently interact in real-world social systems. For instance, the study of social contagions in the health domain has shown competitive dynamics among positive and negative health practices, e.g. smoking vs. jogging (Kuhlman et al. 2011a, 2011b). Most interestingly, health research has uncovered ecological interactions among contagions at different scales, such as the interaction between complex contagions (e.g. health-related attitudes and lifestyle choices), and the spread of simple contagions (e.g. biological pathogens) (Campbell and Salathe 2013).

## 4. New Directions

Recent work on complex contagions points to three main directions for future development. The first concerns the study of how multiple contagions interact within the same network and across networks, in what may be called an ecology of complex contagions. The second concerns the study of how the structure of thresholds and their behavioral consequences can vary by social context. The third area concerns the interaction of diversity and homophily in the spread of complex contagions, where diversity can refer to either the diversity of demographic profiles among one's local peers, or to the broader structural diversity that local peers may be situated within.

### 4.1. Ecologies of Complex Contagions.

Past theoretical research has made significant progress in mapping the behavior of complex contagions within a range of network topologies. Newer work has begun to explore the complexities that arise when multiple kinds of contagions interact in the same network (Su et al. 2016). Moreover, while the content of a contagion undoubtedly influences the spread and interaction of competing behaviors, it may have an impact on network structure as well. An important area of future research concerns how complex contagions shape network structure and how network structure shapes complex contagions, as part of a co-evolutionary process of network formation (Teng et al. 2012).

The process of modeling ecologies of contagions goes hand in hand with a growing effort to model complex contagions in in several new domains of collective behavior. Among the most recent applications is the examination of complex contagions in swarm behaviors. One study showed that complex contagions provided the most robust model of escape reflexes in schools of golden shriner fish, where frightened individuals trigger cascades of escape responses on the basis of a fractional threshold among multiple peers (Rosenthal et al. 2015). Another direction for application concerns the role of complex contagions in cognitive science. Simulation results suggest that complex contagions may be able to account for the emergence and spread of new categories, at the level of both perception and language, consistent with the longstanding view that cultural artifacts depend on principles of emergence and diffusion (Dimaggio 1997; Puglisi et al. 2008; Centola and Baronchelli 2015). Related extensions concern the role of contagion in the structuring of collective memory (Coman et al. 2016). Situating complex contagions at this level will extend existing perspectives on how processes of social diffusion are woven into the foundations of culture and cognition.

*4.2. Mapping Heterogeneous Thresholds in Context.*
Extant models represent threshold heterogeneity in terms of distributions of values along a numerically defined scale, from 0 to 1 (Morris 2000). Applied studies of contagion dynamics show how thresholds vary by individual differences and contextual dependencies relating to the content of the contagion and its sociological significance. For instance, there appear to be a different set of thresholds that govern the adoption of a contagion (e.g. a technology) and the termination of the contagion (Karsai et al. 2014). Similarly, the study of health contagions suggests that people are susceptible to influence by those supporting a positive health behavior and to those resisting it, where individuals may vary in their responses to processes of support and resistance (Kuhlmann et al. 2011a; Myneni et al. 2015).

In the context of social media, readymade sharing buttons alter the cost structure for certain contagions, allowing memes to be adopted simply with a click. Interface design can also make certain contagions more costly, thereby impacting the thresholds of individuals and their willingness to adopt. Certain complex contagions, such as political hashtags, appear to require exposure from 2-5 peers (Barash and Kelly 2012), whereas changes in profile pictures appear to require exposure of up to 8 or more peers (State and Adamic 2015). One conjecture is that thresholds are fractional, and therefore depend sensitively on the number of connections that a person has. The more connections there are, the higher the thresholds are likely to be.

Finally, identity appears to play a structural role in defining thresholds. Identity has been used in two ways: group identity and personal identity. A few recent studies have excluded group identity and focused narrowly on personal identity, such as demographic characteristics (Aral et al. 2009; Aral and Nicolaides 2017). However, the role of demographic characteristics such as gender and race on adoption thresholds is hard to understand independently of social context. Depending on the social context and the identities that are activated, people will react differently to a political contagion than to a health contagion. By contrast, other work has suggested that demographic traits play an important role in defining group identity, which in turn interacts with people's thresholds for adoption (Centola 2011).

Political studies further show how identity-based responses to contagions can take a variety of forms, where thresholds do not simply represent the binary outcome of adoption – they also represent whether an individual will join a committed minority, or whether they will actively attempt to punish deviant behavior (Centola 2013b). Parkinson (2013) uses ethnographic methods to suggest that part of the reason why identity influences contagion thresholds is because identities correspond to different functional roles in a social system, which entail different kinds of behavioral responses that mediate diffusion. These studies help to expose how group membership and normative pressures give rise to individual variation in threshold dynamics. It is likely that individuals differ in the kinds of thresholds they adopt toward a potential contagion based on how they categorize the contagion, relative to their political identity (State and Adamic 2015; Traag 2016). It may therefore be useful to consider different types of thresholds that vary along sociological and psychological dimensions, where key differences are marked by how contagions interface with the identity-based responses of individuals and groups.

*4.3. The Roles of Homophily and Diversity in Diffusion.*
There are two forms of diversity in the literature on diffusion. Re-searchers use the term to refer to cases where one's local neighborhood in the network consists of people with different demographic profiles and personality traits. We may call this *identity-based* diversity. At other times, researchers use the term to refer to *structural diversity* where one's local neighborhood consists of people who belong to separate components of the network, identified by removing the ego node from the ego network. The first kind of diversity tends to limit diffusion of complex contagions, while the second kind tends to amplify it.

Looking at identity-based diversity, Centola (2011) compared complex contagion dynamics on homophilous networks to the dynamics on non-homophilous networks, keeping network topology constant. The results showed that homophily (i.e., reduced identity-diversity) significantly improved the spread of complex contagions. The reason for this is that greater similarity among contacts in a health context made peers more relevant. Women were more likely to adopt from women, and obese people were likely to adopt from obese people. Reinforcing signals from irrelevant (i.e., diverse) peers were largely ignored, while reinforcing signals from relevant (i.e., similar) peers were influential in getting individuals to adopt a new health behavior. This result was most striking for the obese individuals. Exposure among obese individuals was the same across conditions, yet there was not a single obese adopter in any of the diverse networks, while the number of obese adopters in homophilous networks was equivalent to the total number of overall adopters in the diverse networks –resulting in a 200% increase in overall adoption as a result of similarity among peers.

The effects of homophily can be complemented by structural diversity. In studying the complexity of campaign donations, Traag (2016) suggests that structural diversity can increase the credibility of a complex contagion. If one belongs to an echo chamber, where one's peers are highly similar and densely connected, then peer agreement may undermine credibility, since their agreement may be the result of induced homophily and pressures for conformity. By contrast, if one's peers are from different components of the network, their opinions may reasonably be viewed as independent and mutually confirming. What unites these arguments is the supposition that people use the identity composition of their local network neighbors to infer the broader structural diversity of their network. However, structural diversity does not imply reduced homophily. Individuals may be similar to their friends in different ways. They may be the same gender as some, have the same professional role as others, and participate in the same volunteer organizations as yet others. While identity diversity can correlate with structural diversity, it does not always provide a reliable way for inferring it. Receiving reinforcing encouragement from individuals who belong to different parts of a person's social network strengthens the independence of their signals, and may therefore be more likely to trigger adoption.

Similarly, Ugander et al. (2012) identify how the mechanism of structural diversity can boost the influence of social reinforcement. Their study of Facebook shows that people are more likely to adopt a social media technology when they receive invites from people belonging to separate components of their ego network. Structural diversity does not, however, entail identity-based diversity. Ugander et al.'s study leaves open the possibility that structural diversity alone – without identity-based diversity – can modulate adoption thresholds.

This observation is especially interesting in light of State and Adamic's (2015) finding that while the number of friends a user had scaled linearly with their chances of adoption, adoption probabilities plummeted as soon as a user possessed 400 friends or more. The authors propose that having too many friends on social media can stifle the spread of complex contagions by exposing users to a variety of content so vast that they fail to receive repeated exposure by different peers to any given phenomenon. Consistent with earlier results on political hashtags and social movement mobilization, these findings suggest that more contentious complex contagions tend to benefit from clustered, homophilous networks that can foster social change without being overwhelmed by countervailing influences.

## 5. Conclusion

Complex contagions are found in every domain of social behavior, online and off. Early theoretical developments in complex contagions showed that topology and the distribution of adoption thresholds can be decisive for determining whether global saturation is possible. More recent theoretical modeling concerns the interaction of multiple different contagions in the same network, where individuals are attributed different motivations and behavioral responses to each contagion. One of the critical challenges ahead involves mapping heterogeneous thresholds in context, where political identity, group membership, and even the content of contagions can affect individual thresholds and, by consequence, diffusion. Another valuable area for future research concerns the ways in which individuals use information about global network structure to inform their adoption patterns, as is demonstrated by the effects of structural diversity on diffusion. Investigations in this direction will benefit from studying how individuals infer global structure from local interactions, and how new social media environments are augmenting these inferences by supplying information about one's broader ego network. As shown by the literature accumulated over the last decade, examining complex contagions in various applied domains has been enormously fruitful. Each new domain has revealed new elements of diffusion dynamics that require new theoretical explanations and elaborated modelling techniques, revealing new areas of cumulative progress in understanding the collective dynamics of social diffusion.